\documentclass[aps,prb,twocolumn,showpacs,superscriptaddress]{revtex4}

\usepackage{graphicx}

\begin{document}


\title{Softening of the Bond-Stretching Phonon of Ba$_{1-x}$K$_x$BiO$_3$ 
at the Metal to Insulator Transition}

\author{H. Khosroabadi}
\altaffiliation[Present address: ]
{Department of Physics, Sharif University of Technology, Tehran, Iran}
\affiliation{Department of Physics, Graduate School of Science, Osaka University, Osaka 560-0043, Japan}

\author{S. Miyasaka}
\affiliation{Department of Physics, Graduate School of Science, Osaka University, Osaka 560-0043, Japan}
\affiliation{JST, Transformative Research-Project on Iron Pnictides (TRIP), Tokyo 102-0075, Japan}

\author{J. Kobayashi}
\affiliation{Department of Physics, Graduate School of Science, Osaka University, Osaka 560-0043, Japan}

\author{K. Tanaka}
\affiliation{Department of Physics, Graduate School of Science, Osaka University, Osaka 560-0043, Japan}
\affiliation{JST, Transformative Research-Project on Iron Pnictides (TRIP), Tokyo 102-0075, Japan}

\author{H. Uchiyama}
\affiliation{Research and Utilization Division, SPring-8/JASRI, Hyogo 679-5198, Japan}
\affiliation{Materials Dynamics Laboratory, RIKEN SPring-8 Center, Hyogo 679-5148, Japan}

\author{A. Q. R. Baron}
\affiliation{Research and Utilization Division, SPring-8/JASRI, Hyogo 679-5198, Japan}
\affiliation{Materials Dynamics Laboratory, RIKEN SPring-8 Center, Hyogo 679-5148, Japan}
\affiliation{JST, Transformative Research-Project on Iron Pnictides (TRIP), Tokyo 102-0075, Japan}

\author{S. Tajima}
\affiliation{Department of Physics, Graduate School of Science, Osaka University, Osaka 560-0043, Japan}
\affiliation{JST, Transformative Research-Project on Iron Pnictides (TRIP), Tokyo 102-0075, Japan}

\date{\today}

\begin{abstract}
The dispersion of phonons in Ba$_{1-x}$K$_x$BiO$_3$ along the (3+$q$, 0, 0) direction in reciprocal space was determined for $x$=0, 0.30, 0.37, 0.52 using high-resolution inelastic X-ray scattering.  The observed phonon energies near $\Gamma$ were in good agreement with published optical data. It was found that two high-energy modes strongly soften near $q$=0.25 when the system becomes metallic ($x>$0.35). There was no softening in the insulating phase, even at $x$=0.30, very near the metal-insulator transition. This suggests that the phonon softening is linked to the appearance of the Fermi-surface. 
\end{abstract}



\maketitle


\section{Introduction}
The electron-phonon interaction plays a central role 
in conventional BCS-Eliashberg superconductors. 
While, in most cases, the phonons contributing to 
superconductivity are low frequency modes, 
there are two examples of superconductors 
in which a high frequency optical mode predominantly contributes to pairing: 
Ba(Pb,Bi)O$_3$ with the superconducting transition temperature, 
$T_{\mathrm c}$=11 K \cite{Sleight1} and (Ba,K)BiO$_3$ 
with $T_{\mathrm c}$=30 K \cite{Cava1}. 
In these materials the Bi-O breathing mode phonon strongly 
couples to the electronic system \cite{Mattheiss0,Shirai1}. 
Another example is MgB$_2$, $T_{\mathrm c}$=39 K 
\cite{Nagamatsu1}, where the coupling to the high-energy 
boron optical mode plays a main role 
in superconductivity \cite{Kortus1}. 

Strong electron-phonon coupling often manifests itself 
as phonon softening. 
For MgB$_2$, inelastic X-ray scattering (IXS) has 
demonstrated that the dispersion of the $E_{2g}$ phonon 
strongly softens and broadens 
near zone center \cite{Shulka,Baron1}. 
Subsequent work with carbon substituted MgB$_2$ revealed 
that the amount of phonon softening correlates 
with $T_{\mathrm c}$ \cite{Baron2}, providing direct proof 
that this phonon is responsible for the high $T_{\mathrm c}$ 
superconductivity in MgB$_2$. 

By contrast, in the high-$T_{\mathrm c}$ superconducting 
cuprates (HTSCs), the phonon contribution to superconductivity 
remains unclear, although the high-frequency oxygen mode 
(Cu-O bond stretching mode) shows strong softening 
as the momentum transfer is 
changed\cite{Pintschovius1,McQueeney1,Astuto1,Uchiyama1,Pintschovius2,
Fukuda1,Reznik1,Graf1}. 
The main subject of discussion for the cuprates is the rapid drop 
of the bond stretching phonon frequency, and increased in linewidth, 
in the ($q$, 0, 0) direction at $q>$ 0.25, 
which is not reproduced by the LDA 
calculations\cite{Pintschovius2,Reznik2,Bohnen}. 
The carrier doping dependence of the phonon softening is also
an unresolved problem. 
Namely, why the softening gradually develops with hole doping and saturates 
at the high doping levels \cite{Fukuda1} is not well understood. 

The latter problem motivated us to investigate the doping dependence of
phonon softening in other superconducting systems near a metal-insulator (MI)
transition.
Phonon softening near a MI transition is usually observed when temperature
is lowered.
However, there are few studies of phonon softening as a function of carrier doping
by chemical substitution, 
particularly in system where there is a doping-induced MI transition.
From this viewpoint, (Ba,K)BiO$_3$ (BKBO) is a good candidate for study.
BKBO has a perovskite structure,
similar to the cuprates.
As K is doped into the system, BKBO changes from 
a charge-density-wave (CDW) insulator to a metal 
that superconducts ($T_{\mathrm c} \sim $ 30K) at $x \sim$ 0.35. 
Crystal structure also changes with $x$ from monoclinic 
to orthorhombic (at $x \sim $ 0.1) and then to cubic structure 
(at $x \sim$ 0.35) \cite{Pei1}. 

The phonon dispersion of BKBO has previously been 
investigated only by neutron scattering measurements 
for K concentrations of $x$=0.02 and 0.40 \cite{Braden1,Braden2}. 
That work showed that no phonon softening was observed 
in the insulator, while dramatic softening of the Bi-O stretching 
and breathing modes appeared in the metallic (superconducting) phase. 
It is then interesting to consider how the softening changes 
as one increases $x$ through the MI phase 
transition, for example, whether the phonon softens only 
near the MI transition or whether the phonon softening 
correlates with $T_{\mathrm c}$ as the electron-phonon 
coupling changes with K concentration. 
However, the earlier work \cite{Braden1,Braden2} was limited 
to only the two K concentrations. 

In the present paper we investigate the phonon dispersion of 
BKBO single crystals with $x$=0, 0.30, 0.37, and 0.52 
in order to clarify the phonon behavior near the MI transition. 
Despite great effort, we were not able to grow single-phase 
samples with dimension larger than about 0.5 mm, however, 
our technique of measuring phonon dispersion, 
IXS, allows us to perform 
these measurements on such small samples. 
The samples with $x$=0 and 0.30 are in the insulating phase, 
while those with $x$=0.37 and 0.52 are metallic. 
So far, there has been no report of phonon dispersion 
for $x$=0.30, near the MI transition, and for $x$=0.52, 
a heavily doped composition of the metallic phase. 

\section{Experiments}
Single crystals of BaBiO$_3$ were grown by a flux method 
using Bi$_2$O$_3$ and BaCO$_3$, and the K-doped crystals 
by the electrochemical technique using Bi$_2$O$_3$, 
Ba(OH)$_2$$\cdot$8H$_2$O and KOH 
\cite{Norton1,Nishio1,Khosroabadi1}. 
Our sample sizes were typically $\sim $ 0.3$\times$0.3$\times$0.3 mm$^3$ 
($<$0.2 mg) - while effort was made to find larger crystals, 
they were always multi-domain and/or non-uniform in doping. 
It was particularly difficult to grow homogenous samples with $x \sim$0.3, 
and this may explain why there are relatively few experimental 
investigations of samples with $x \sim$0.3 to 0.35. 
The lattice constants of the samples were measured 
using a four-circle X-ray diffractometer, and the K concentration 
was determined assuming a linear relationship 
between the lattice parameter and $x$ in Ba$_{1-x}$K$_x$BiO$_3$ \cite{Pei1}. 
Superconducting transition temperatures were determined 
from the magnetic susceptibility measured by a superconductor quantum 
interference device (SQUID) magnetometer. 
The K-content estimated from the lattice parameter was 
in good agreement with the value expected 
from the measured $T_{\mathrm c}$ \cite{Pei1}. 

IXS spectra were measured 
at BL35XU of SPring-8 \cite{Baron3}. 
We measured spectra along (100) direction of the simple 
perovskite lattice, i.e. from $\Gamma$ to X points, (3+$q$,0,0), 
where 0$<q<$0.5, at room temperature. 
Depending on setup, the energy resolution varied from 
1.5 to 6 meV according to the choice of x-ray energy 
and analyzer Bragg reflection (1.5/3/6 meV resolution 
at the Si (11 11 11)/(999)/(888) at 21.7/17.8/15.8 keV). 
The incident beam at the sample position was about 60$\times$ 
100 $\mu$m$^2$ in the full-width at half maximum (FWHM). 
Because of the strong X-ray absorption by heavy elements, 
the X-ray penetration depth in BKBO is small ($<$20 $\mu$m), 
for example, about 40$\%$ of that for (La,Sr)$_2$CuO$_4$ 
which resulted in a relatively weak phonon signal. 
For the analysis of the IXS spectra, the pseudo-Voigt 
functions were used. 

 \begin{figure}[htp]
 \includegraphics[width=50mm]{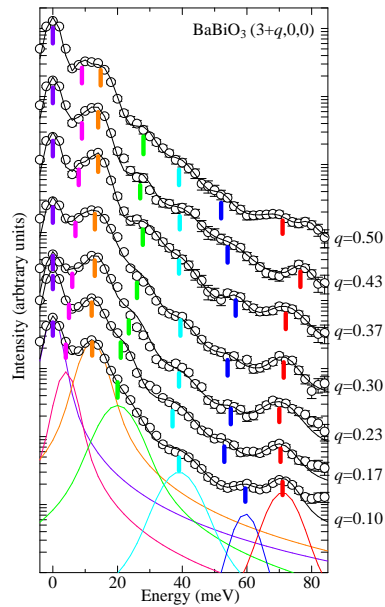}
 \caption{(Color online) A typical IXS spectrum for BaBiO$_3$ system 
at various $q$ values. 
The open circles show the experimental data with their statistical 
error bars. 
The lines below the spectrum show the pseudo-Voigt fitting 
for elastic scattering and each phonon mode indicated by a short bar. 
The black solid lines show the total fitting of the experimental data.}
 \label{fig1}
 \end{figure}

\section{Results}
Figure 1 shows the IXS spectrum of BaBiO$_3$ at various $q$ values. 
The spectra are well reproduced by fitting curves consisting of 
an elastic scattering peak and six phonon peaks. 
Since the measurement direction is (3+$q$, 0, 0) 
in the present experiment, the intensity of longitudinal optical 
(LO) phonons is enhanced, while the transverse optical (TO) 
phonons are almost undetectable. 
Since the intensity at energy higher than 80 meV is very small, 
we consider the features in this energy range unreliable 
and ignore them. 
Based on the published optical data \cite{Uchida1,Sugai1}, 
we assign the six peaks from low to high frequency 
as the acoustic mode, external mode (relative displacement 
of Ba and BiO octahedron), Bi mode, Bi-O bending mode, 
Bi-O stretching mode, and Bi-O breathing mode. 
Although there are only four longitudinal modes, 
including the acoustic mode, 
for a cubic perovskite structure, the lattice distortion 
coupled with CDW creates additional phonon branches. 
For example, the Bi-O breathing mode at the zone boundary 
becomes a zone center mode that is Raman active. 
The Bi-Bi vibration mode that is originally an acoustic mode 
turns to an infrared active mode because of inequivalent 
neighboring Bi charges \cite{Uchida1}. 

 \begin{figure}[htp]
 \includegraphics[width=60mm]{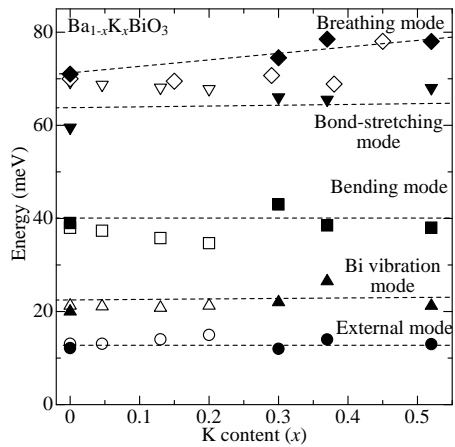}
 \caption{The potassium content dependence 
of phonon energies for various modes at $q \sim $0.1 
in (3+$q$,0,0) momentum (closed symbols). 
Open symbols are the other experimental data for the LO phonon modes 
from Raman and optical spectroscopies 
\cite{Sugai1,McCarty1,Tajima1}. 
The dashed lines are guide for eyes.}
 \label{fig2}
 \end{figure}

Figure 2 illustrates a comparison of our data near the $\Gamma$ 
point ($q \sim$0.1) with published optical data 
including both far-infrared and Raman scattering 
measurements \cite{Uchida1,Sugai1,McCarty1,Tajima1,Nishio2}. 
As seen in the figure, our IXS data agree well 
with the optical data over the whole composition range, 
except for the peak at 60 meV. This is strong support 
for the reliability of our measurement. 
As for the 60 meV peak, we tentatively attribute it 
to the LO bond-stretching mode, although its energy is different 
from the far-infrared data (69 meV). 
Another possibility might be a TO-mode, which, 
while it should be weak in our IXS data, 
is known to be at 60 meV from far-infrared measurements.

 \begin{figure}[htp]
 \includegraphics[width=60mm]{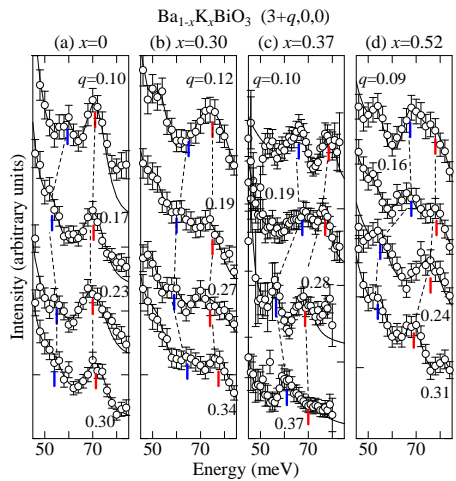}
 \caption{(Color online) High energy part of IXS spectrum 
of BKBO system for the four compositions. 
The number beside the data shows the $q$ value in (3+$q$,0,0) momentum. 
The open symbols show the experimental data and the solid lines 
show the fitting result. 
The short bars on the data indicate the positions of the phonon peaks. 
The vertical dashed lines are guide for eyes to track 
the phonon dispersions. 
The data were plotted by a shift in the vertical axis 
to distinguish the different $q$ patterns.}
 \label{fig3}
 \end{figure}

We now focus on the higher frequency phonons. 
In Fig. 3, the high-energy part of the spectra ($\hbar \omega >$
45 meV) is shown at several $q$-values for all of the measured crystals. 
For the insulating samples with $x$=0 and 0.30, the $q$-dependence 
of the phonon frequency is weak, while the phonons 
in the metallic samples with $x$=0.37 and 0.52 exhibit 
large changes between the second and 
the third $q$ vectors (at $q \sim$0.2). 

The observed phonon frequencies at all the $q$-values 
are summarized in Fig.4. 
Anomalous softening of the bond stretching and the breathing modes 
is observed for $x$=0.37 and 0.52, while almost no softening 
is seen for $x$=0 and 0.30. 
The neutron scattering results of Braden \cite{Braden1,Braden2} 
are also plotted for comparison. 
The agreement with the neutron data is good for $x$=0, 
but not for $x$=0.37. 
This discrepancy between our data and the neutron result 
might be due to a difference in K-content of the samples. 
Our IXS results are in better agreement with the optical data. 
Although the phonon frequencies are different, 
the tendency of phonon softening is the same in both neutron 
scattering and IXS experiments. 

 \begin{figure}[htp]
 \includegraphics[width=60mm]{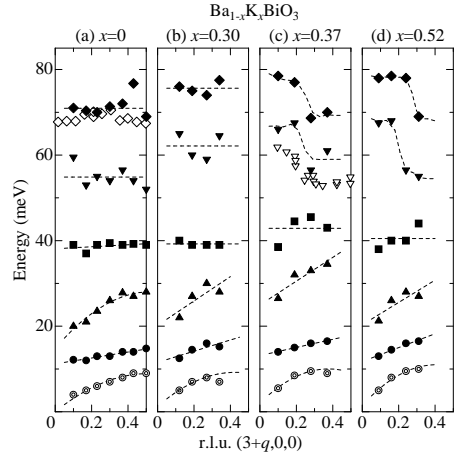}
 \caption{Phonon branches of BKBO system in (100) direction 
for four different values of $x$ (closed symbols and double circles) 
and the data from neutron scattering study 
\cite{Braden1,Braden2} for comparison (open symbols). 
The dashed lines are guides for eyes.}
 \label{fig4}
 \end{figure}

It should be noted that the breathing mode as well as 
the Bi-Bi vibration mode survives even in the metallic phase, 
although these modes should not be present in a cubic perovskite 
structure without the CDW distortion. 
The present results suggest that local lattice distortions remain 
even in the nominally cubic phase. 
In fact, we observed strong superlattice peaks 
during the experiments, as if the CDW distortion remained. 
Also there have been several reports on the short range 
structure distortion suggested 
by X-ray absorption fine structure \cite{Yacoby1}, 
ultrasonic measurements \cite{Zherlitsyn1} 
and Raman scattering \cite{Pashkevich1} in the metallic phase. 
Recently, even a neutron diffraction measurement 
has detected a long-range structural distortion 
in the superconducting BKBO characterized 
by a tilt of the BiO octahedra \cite{Braden3}. 

\section{Discussions}
An important result of the present study is that no softening 
is observed in the phonon dispersion for $x$=0.3, 
very close to the MI transition in the insulator phase. 
Often (in a conventional CDW material, for example), 
phonon softening occurs as the temperature is swept 
through the metal-insulator transition, 
with the softening being largest at the phase transition, 
and then recovering. 
In the present case, we change the doping 
at a fixed temperature (300K), and the crystal structure changes 
from orthorhombic to cubic at $x \sim$0.35, 
accompanied by the change from a CDW insulator to a metal. 
(The change from monoclinic to orthorhombic at $x$=0.1 
does not seem to affect the phonon dispersion.) 
Making an analogy between $T$- and $x$-sweep experiments, 
we expected phonon softening at $x$ just before 
and after the MI transition. 
The absence of the phonon softening at $x$=0.3 may be 
because the MI transition is too sharp as a function of $x$ (or $T$) 
to follow the transient behavior of the phonon. 
Since even near $x$=0.35 the MI transition temperature 
is much higher than room temperature, 
we cannot observe softening at 300K. 
This fact implies that the CDW gap is still large at $x$=0.30 
as was suggested optical absorption data 
($\sim $ 0.5eV at $x$=0.28, while 2eV at $x$=0) \cite{Sato1}. 

In contrast, clear phonon softening in BKBO is observed 
in the metallic phase with a large Fermi surface (FS). 
Once the FS is formed after the phase transition, 
it changes only weakly with $x$, 
which might result in the $x$-independent ($x>$0.35) softening 
seen in the present work. 

According to the theoretical prediction, 
the electron-phonon coupling constant, $\lambda$, 
which has the largest contribution from the breathing mode, decreases 
with increasing $x$ \cite{Shirai1}, because the FS shrinks, 
as the band-filling is reduced. 
This can explain why $T_{\mathrm c}$ for $x$=0.52 is lower than 
that for $x$=0.37. 
However, the present results do not show a clear $x$- (or $T_{\mathrm c}$-) 
dependence of phonon softening in the metallic phase. 
Therefore, at least in the present case, the amount 
of softening of the bond stretching phonon is not a good 
measure for the strength of the electron-phonon coupling that determines $T_c$. 

The reason for this unexpected result is not clear. 
One possibility is that the phonon change is too small to be observed with the 
present experimental resolution. 
According to the Shirai's calculation \cite{Shirai1}, 
$\lambda$ changes from 0.88 ($x$=0.4) to 0.53 ($x$=0.5), 
from which $T_c$ was estimated as 25K ($x$=0.4) and 10K ($x$=0.5), 
on the basis of the Allen-Dynes equation\cite{Allen1}. 
These values are not far from the experimental data 
(30 K for $x$=0.4 and 17 K for $x$=0.5)\cite{Pei1}. 
Compared to such a drastic change in $T_c$ with $\lambda$, 
the phonon frequency is expected to show a weaker $\lambda$-dependence. 
The phonon renormalization due to electron-phonon coupling is generalized as 
$\omega_{ph}^2 \sim \Omega_{ph}^2/(1+\lambda)$ \cite{Pickett1}, 
where $\omega_{ph}$ is a renormalized phonon frequency 
and $\Omega_{ph}$ is a bare phonon frequency\cite{comment1}. 
This expression gives about 10 \% change in phonon frequency 
when $\lambda$ changes from 0.88 ($x$=0.4) to 0.53 ($x$=0.5), 
which is much smaller than the 60 \% reduction in $T_c$. 
While the above relation between $\lambda$ and the renormalized phonon frequency 
is a very rough approximation, 
the important idea in the present context is that the fractional changes 
in phonon frequency may be small compared to changes in $\lambda$.

Finally, we note that the present result brings the unusual 
phonon behavior of the HTSCs into sharp relief. 
Softening of the bond stretching phonon 
has been reported in cuprates with the layered perovskite structure. \cite{Pintschovius1,McQueeney1,Astuto1,Uchiyama1,Pintschovius2,Fukuda1,Reznik1,Graf1},
however, the doping dependence in the cuprates\cite{Fukuda1} is very different.
In BKBO, the electronic change with doping is small in the metallic phase, 
where a rigid band picture is applicable as a first approximation. 
In this situation, as we see in the present study, even near the MI transition, 
10 \% change of band filling \cite{comment2} gives 
only a little change in phonon softening.
By contrast, in the underdoped cuprates 
(for example, at $0<x<0.15$ of La$_{2-x}$Sr$_x$CuO$_4$), 
the system is in the so-called "pseudogap" state \cite{Timusk} 
where the FS rapidly develops with carrier doping
from a Mott insulator \cite{Norman1}. 
Therefore, it is likely that such a radical change of FS topology accompanied with a huge enhancement of the electronic density of states is responsible 
for the gradual increase in softening of the Cu-O bond stretching phonon with carrier doping. The importance of FS topology in electron-phonon coupling of the cuprates 
was also pointed out in ref. \cite{Graf1}. 
Further detailed information about FS will promote more precise calculations of the phonon behaviors in the cuprates.

\section{Summary}
We have investigated the phonon dispersion of 
Ba$_{1-x}$K$_x$BiO$_3$ in the (3+$q$, 0, 0) direction)
for $x$=0, 0.30, 0.37, and 0.52 using IXS. 
In contrast to the nearly doping independent behavior 
of the low frequency modes, the higher frequency modes related 
to Bi-O bond stretching vibrations showed a clear 
and abrupt change with K-content. 
In the insulating phase, $x<$0.35, mode dispersions 
were almost flat, while radical softening was observed 
towards zone boundary in the metallic phase, $x>$0.35. 
The degree of softening is relatively unaffected by K-content 
($x>$0.35). 
This doping dependence of the phonon softening suggests 
that the present softening is different than the softening 
often associated with a structural phase transition, 
and, more probably, is electronic in origin.  

\begin{acknowledgments}
The authors thank T. Oguchi for a useful discussion. 
This work was partially supported by the Grant-in-Aid for Scientific 
Research (Grant No. 19204038) from the Ministry of Education, 
Culture, Sports, Science and Technology of Japan. 
H. K. appreciates the support by the Japan Society 
for the Promotion of Science (JSPS) scholarship 
and its research grants. 
\end{acknowledgments}


\end{document}